\documentstyle[11pt,psfig]{article}
\def\be{\begin{equation}}
\def\en{\end{equation}}

\begin{document}
\baselineskip = 24pt

\Large

\begin{center}
{\bf Non-axisymmetric relativistic Bondi-Hoyle accretion
       onto a Schwarzschild black hole}
\end{center}

\normalsize

\vspace{0.5cm}

\begin{center}
Jos\'e A. Font$^{1}$ and
J.M$^{\underline{\mbox{a}}}$ Ib\'a\~nez$^{2}$

\vspace{0.2cm}

$^{1}$
Max-Planck-Institut f\"ur Gravitationsphysik,  
Albert-Einstein-Institut \\
Schlaatzweg 1, 14473 Potsdam, Germany \\
e-mail: font@aei-potsdam.mpg.de

$^{2}$
Departamento de Astronom\'{\i}a y Astrof\'{\i}sica\\
Universidad de Valencia, 46100 Burjassot (Valencia), Spain \\
e-mail: ibanez@scry.daa.uv.es

\end{center}

\vspace{1cm}

\begin{abstract}

\noindent

We present the results of an exhaustive numerical study of
fully relativistic non-axisymmetric Bondi-Hoyle accretion onto
a moving Schwarzschild black hole.
We have solved the equations of general relativistic
hydrodynamics with a high-resolution
shock-capturing numerical scheme based on a linearized
Riemann solver. The numerical code was previously used to study axisymmetric
flow configurations past a Schwarzschild hole.
We have analyzed and discussed the flow morphology for a sample of
asymptotically high Mach number models. 
The results of this work reveal that initially asymptotic uniform
flows always accrete onto the hole in a stationary way
which closely resembles the previous axisymmetric patterns.
This is in contrast with some Newtonian numerical studies where
violent flip-flop instabilities were found.
As discussed in the text,
the reason can be found in the initial conditions used in the
relativistic regime, as they can not exactly duplicate
the previous Newtonian setups where the instability appeared.
The dependence of the final solution with the inner
boundary condition as well as with the grid resolution has also been studied.
Finally, we have computed the accretion rates of mass and linear and angular
momentum.
\end{abstract}


\newpage

\section{Introduction}

In a previous paper (Font and Ib\'a\~nez 1997, hereafter Paper I) we studied
the morphology and dynamics of relativistic Bondi-Hoyle accretion in
axisymmetric flows past a Schwarzschild black hole. The main conclusion
of that work was to extend the validity of the Bondi-Hoyle
accretion picture (Hoyle and Lyttleton, 1939; Bondi and Hoyle, 1944)
to the relativistic regime, finding that the matter is
always accreted onto the hole
in a stationary way. 
Furthermore, if the flow was, initially, supersonic, the main
feature of the accretion pattern was the presence of a shock cone in
the solution.
At the same time, we checked the
validity of our numerical code revisiting an existing 
previous calculation (Petrich et al., 1989) using more accurate numerical
techniques specifically designed to capture discontinuities.
In the present investigation we have extended those studies to 
account for non-axisymmetric configurations. 
We have only considered uniform flows at infinity
focussing on studying whether or not the flow pattern reaches,
ultimately, a stationary state. 
The main purpose of this work is, then, to
extend previous non-axisymmetric Newtonian computations to the
realm of general relativity. As far as we know, this has not been
studied so far. 
In particular, we plan to address if non-steady patterns, so 
prominent in some of the 2D non-axisymmetric simulations in the
classical regime, also arise here.

During the last years, a large number of wind accretion 
simulations past a gravitating body have been performed in Newtonian
hydrodynamics (see, e.g., the list of references in Paper I).
One of the most interesting features, only revealed by numerical
simulations, was the appearance of
unstable accretion patterns, contrary to the theoretical 
(and simplified) Bondi-Hoyle (1944) accretion picture. 
This highly non-steady behaviour on the wake of the accretor
was only found in 2D non-axisymmetric simulations, especially when the
resolution employed was fine enough. 
These unstable patterns are characterized
by the shock wave moving from side to side (the so-called flip-flop
 instability) and also by the 
appearance of transient phases of disc formation where the angular
 momentum accreted by the central object significantly increases.
This kind of behaviour has been found not only assuming local density
and velocity gradients (Fryxell and Taam, 1988; Taam and Fryxell, 1989)
but also when considering accretion of uniform
flows at infinity (Matsuda et al., 1991).
However, in detailed 3D computations performed more recently 
(Ruffert and Arnett, 1994; Ruffert, 1994a,b, 1995, 1996) the accretion
cone remains quite stable and no sign of flip-flop instability appears unless
the flow is assumed to have explicit density gradients at infinity
(Ruffert, 1997).

In all previous studies it was shown that the key parameter which
controls the appearance of the instability was the 
{\it size of the accretor}. In particular,
this size should be a very small fraction of the accretion radius in order to
find any evidence of flip-flop. For
instance, in Sawada et al. (1989) and Matsuda et al. (1991) computations,
the flip-flop instability appeared, in 2D, for a central size object of
$\approx 0.0625 r_a$, where $r_a$ is the accretion radius, the natural
length scale of this problem, defined as

\begin{equation}
r_{a} = {{2GM}\over {v_{\infty}^{2}}}. 
\end{equation}

\noindent
Here, $M$ is the mass of the accreting object, $G$ is the
gravitational constant and $v_{\infty}$ is the asymptotic velocity of the
fluid. They also found that the flow was stable if the size was larger
than $\approx 0.125 r_a$. More recently, 
Benensohn, Lamb and Taam (1997) have performed
detailed 2D non-axisymmetric computations with a smaller accretor of size
$0.0375 r_a$ finding unstable behaviour as well.

In addition, there are other parameters that, combined with the size
of the accretor, can play
a role in the development of the instability. In particular, one of these
is the {\it Mach number} of the flow. 
The larger the Mach number at infinity
is, the more turbulent the wake becomes.
Finally, another extremely
important consideration is the {\it position of the shock},
namely if it is attached (tail shock) or detached (bow shock) 
with respect to the
central object. This is basically controlled by the values of the
asymptotic Mach number and the adiabatic index of the gas. For low
Mach number and high adiabatic index flows (say $5/3$), the shock
is mainly detached at the front while the opposite trend produces
a tail shock.
All previous studies where the
instability appeared needed the formation of
a detached shock in front of the accretor. As the evolution procceeded,
this detached shock transformed into a dome-shaped shock and,
eventually, this was followed by a violent flip-flop instability
(Sawada et al., 1989; Matsuda et al., 1991; Livio et al., 1991;
see also Ruffert, 1997). In particular, Livio et al. (1991) demonstrated,
using a simplified analytic treatment, the existence of a wake instability
to small deflections when the shock opening angle is large. This can 
happen even for an initially homogeneous flow if the accretion process has
low efficiency. However, the flow can be stabilized if the accreting
object is larger than some critical dimension (Livio et al., 1991).

According to these previous classical results, there are several factors
that can lead to the non-appearance of this instability in the
relativistic regime when the accretor is a 
black hole. First, and most important, 
in this regime there is a physical minimum value for the
size of the accreting object, given by the gravitational
(Schwarzschild) radius. 
This, of course, places some constraints
in the possibility of extending previous Newtonian studies to very small
values of the accretor, assuming that we also need an asymptotically
supersonic flow (in order to have any shock wave) and avoiding, at the
same time, flow velocities larger than the speed of light. 
The second factor has to do with the position of the shock. As we will show
below (see also Paper I), all models that we have evolved are characterized
by the presence of a {\it tail} shock in the rear part of the hole.
Hence, according to the Newtonian studies, this would not help the
development of the flip-flop instability.

The paper is organized as follows: in next Section ($\S 2$)
we briefly present the system of equations of general relativistic
hydrodynamics written as a hyperbolic system of
conservation laws. The numerical code and the initial setup
are also described here.
The results of the simulations are presented and analyzed in
Section $\S 3$.
Finally, Section $\S 4$ summarizes the main conclusions of this
work.

\section{Equations, initial setup and numerical issues}

\subsection{Equations}

In order to study non-axisymmetric patterns in two dimensions we have
restricted ourselves to an infinitesimally thin disk in the equatorial
plane of the black hole. Therefore, we are using ($r,\phi$) coordinates,
instead of ($r,\theta$) used previously in Paper I.
Although this configuration is somehow artificial,
it suffices to try to understand the stability of the flow. This kind
of setup has also been used in Newtonian simulations, assuming
flow past an infinite cylinder (Fryxell and Taam, 1988;
Taam and Fryxell, 1989; Benensohn et al., 1997).

With the same definitions introduced in Paper I and using
geometrized units ($G=c=1$ with $c$ the speed of light), 
the equations of (adiabatic) general relativistic hydrodynamics can
be written, in the equatorial plane ($\theta=\pi/2$) of the
Schwarzshild metric, as

\begin{equation}
 {{\partial {\bf U}({\bf w})} \over {\partial t}} +
 {{\partial {\bf F}^{r}({\bf w})} \over {\partial r}} +
 {{\partial {\bf F}^{\phi}({\bf w})} \over {\partial \phi}}
 = {\bf S}({\bf w}).
\end{equation}
\noindent
In this equation, the vector of {\it primitive variables} is
defined as
\begin{equation}
{\bf w} = (\rho, v_{r}, v_{\phi}, \varepsilon)
\end{equation}
\noindent
where $\rho$ and $\varepsilon$ are, respectively, the rest-mass density
and the specific internal energy, related to the pressure
$p$ via an equation of state which we chose that of an ideal gas law
\begin{equation}
p=(\gamma - 1) \rho \epsilon.
\end{equation}
\noindent
with $\gamma$ being the constant adiabatic index.
In addition, $v_{r}$ and $v_{\phi}$ denote the radial and azimuthal 
covariant components of the velocity. On the other hand, 
the vector of unknowns (evolved quantities) in equation (2) is
\begin{equation}
{\bf U}({\bf w})  =   (D, S_r, S_{\phi}, \tau).
\end{equation}
\noindent
The explicit relations between both sets of variables are
\begin{equation}
D =  \rho W
\end{equation}
\begin{equation}
S_j  =  \rho h W^2 v_j \,\, (j=r,\phi)
\end{equation}
\begin{equation}
\tau =  \rho h W^2 - p - D.
\end{equation}
\noindent
The quantity $W$ stands for the Lorentz factor, which satisfies
$W=(1-{v}^{2})^{-1/2}$ with
${v}^{2}= \gamma_{ij} v^i v^j$, where $v^i$ is the 3-velocity of the
fluid, defined, for the case of a zero shift vector, according to
$v^i= {{u^i}\over{W}}$
and $\gamma_{ij}$ are the spatial components
of the spacetime metric where the fluid evolves. The quantity $h$
appearing in equations (7) and (8) is the specific enthalpy,
defined as $h=1+\epsilon+ p/\rho$.
The corresponding fluxes in the radial and azimuthal
directions are, respectively
\begin{equation}
{\bf F}^{r}({\bf w})  =   \left(D v^{r},
 S_r v^{r} + p, S_{\phi} v^{r},
(\tau + p) v^{r} \right)
\end{equation}
\noindent
\begin{equation}
{\bf F}^{\phi}({\bf w})  =   \left(D v^{\phi},
S_{r} v^{\phi}, S_{\phi} v^{\phi} + p,
(\tau + p) v^{\phi} \right)
\end{equation}
\noindent
and the corresponding vector of sources ${\bf S}({\bf w})$ is
\begin{eqnarray}
{\bf S}({\bf w}) =&
\left(
D v^r \left[{{M}\over{\alpha r^2}} - {{2\alpha}\over{r}}\right],
-{{M}\over{\alpha r^2}}(\tau + D) +
\alpha \left[{{1}\over{r}}S_{\phi}v^{\phi} - 
             {{2}\over{r}}S_r v^r \right],
\right. \nonumber \\ &
S_{\phi}v^r \left[{{M}\over{\alpha r^2}} - {{2\alpha}\over{r}}\right],
\left.
-{{2\alpha}\over{r}} S^r - 
D v^r \left[{{M}\over{\alpha r^2}} - {{2\alpha}\over{r}}\right]
\right) 
\end{eqnarray}
\noindent
where $\alpha = \sqrt{1-{2M\over r}}$.

\subsection{Initial setup}

The models that we have evolved numerically are summarized in 
Table 1. Some of them were already evolved in axisymmetry in
Paper I. Now, we have added some higher Mach number models and we
have not considered subsonic and low Mach number flows. According
to Newtonian studies, this is the appropriate initial condition to
look for wake instabilities. We must also mention that, contrary to the
axisymmetric results, it was now extremely difficult 
to evolve low
Mach number models without corrupting the asymptotic initial 
values in the upstream region of the flow. In this part of the
domain, all fluid variables should increase monotonically.
In practice, we found that
the local velocity and Mach number in the upstream direction fell
below the initial values. This was still the case even when we placed
the outer boundary further out (although the evolution could
be accurately followed for much longer times). 
This sort of behaviour has not
appeared in the high Mach number models, as we
will show in detail below. This inherent difficulty in
approximate the asymptotic conditions to a finite distance has also
been pointed out in previous Newtonian simulations (Ruffert
and Arnett, 1994; Benensohn et al., 1997).

The flow pattern is completely characterized by
some initial conditions at infinity upstream the hole, namely the
velocity $v_{\infty}$, sound speed $c_{s_{\infty}}$
and adiabatic index, $\gamma$. The first two parameters fix the asymptotic
Mach number, ${\cal M}_{\infty}$.
The covariant components of the initial velocity are given in
terms of its asymptotic value:
\begin{equation}
v_r = {{1}\over{\sqrt{\gamma^{rr}}}} v_{\infty} cos\phi
\end{equation}
\begin{equation}
v_{\phi} = - {{1}\over{\sqrt{\gamma^{\phi\phi}}}} v_{\infty} sin\phi
\end{equation}
\noindent
where $\gamma^{rr}=1-{{2M}\over{r}}$ and 
$\gamma^{\phi\phi}={{1}\over{r^2}}$.

As mentioned in the introduction, it is convenient to stress the
differences that our initial setup has with respect to those of classical
simulations where the tangential wake instability arose. 
For this purpose, we plot in
Fig. 1 the ratio between the size of the accretor, $2M$, the Schwarzschild
radius, and the accretion radius, as a function of the asymptotic
flow velocity. For the accretion radius we are using the same
definition of Paper I
\begin{equation}
r_a = {{M}\over{(v^2_{\infty} + c_{s_{\infty}}^2)}}.
\end{equation}
\noindent
The curved lines represent that quotient for different
values of the sound speed at infinity. The solid line corresponds to
zero sound speed and if we move upward on the plot, we go to increasingly
larger values. The last three curves correspond to the 
thermodinamically maximum permitted values,
$\sqrt{\gamma-1}$, for $\gamma=1.1$, $4/3$ and $5/3$, respectively. On the
other hand, the straight lines represent 
the fluid Mach number (divided by 5) as a function
of the velocity, parametrized for the same set of sound speeds. Therefore,
the lines are drawn accordingly with the same
line style, with the exception of the solid straight line 
which corresponds
to $c_{s_{\infty}}=0.05$ (instead of zero sound speed). 
From that figure, one can see that, in order
to have tiny accretors, one has to move to the lower left corner. 
However, in that region,
the asymptotic initial velocity has to be quite small (Newtonian flow).
Furthermore, in order to have an initial
supersonic, or even better, hypersonic flow, the initial sound speed
has to be considerably small.
This kind of initial data is quite difficult to 
evolve for any numerical code, as the internal energy density is
very close to zero.
On the other hand, if we move to larger values of 
the velocity, for the sample of sound speeds considered here, the
size of the accretor has to be huge. 
In the limiting case
of maximum sound speed and maximum velocity, the minimum size of the
accretor is larger than $2r_a$ for $\gamma=1.1$ and $4/3$ and even
larger than $3r_a$ for $\gamma=5/3$.
According to this, and having
in mind the classical results, it would not be too surprising that
the relativistic patterns would appear much more stable. 

Finally, concerning the numerical grid,
the spatial numerical domain, $({r, \phi})$, has been covered by
a canonical grid of $200 \times 80$ numerical zones.
The radial and angular discretizations lie, respectively,
in the interval
$r_{min} \le r \le r_{max}$ and $0 \le \phi \le 2\pi$,
where $r_{min}$ and $r_{max}$ depend on the particular model. The
specific values are given in Table 1.
Note that, with the coordinates we are using, $r_{min}$ will always be
slightly larger than $2M$ in order to avoid divergences 
at the horizon in some
hydrodynamical quantities such as, e.g., the Lorentz factor.
For the angular direction we have used an equally spaced grid while
in the radial direction we have employed the
Schwarzschild {\it tortoise coordinate} defined by
$r_* = r + 2 M \ln ({{r}\over{2M}} -1)$. 

\subsection{Numerical issues}

We have solved the general relativistic
hydrodynamic equations with the same numerical code described in
Paper I. We briefly summarize its main features here.
The code belongs to the so called TVD (Total Variation Diminishing) schemes
(Harten, 1983) and makes use of a linearized
Riemann solver (Roe, 1981) as an accurate procedure to capture
discontinuities. It also uses a monotonic upstream reconstruction
procedure (van Leer, 1979)
to extrapolate the variables at the cell interfaces to
solve for the Riemann problem and a third order Runge-Kutta 
scheme, which preserves the TVD property, 
to update the solution in time. Further details of the code can
be found in Paper I and Font et al. (1994).

The boundary conditions employed in the simulations are, basically, the
same as we used in Paper I. For the innermost radial boundary we have
considered now two different sets of conditions, 
namely {\it outflow}, where all 
variables are linearly extrapolated to the boundary zones, and 
{\it absorbing conditions}, where the velocities are reset to zero and
the density and internal energy are fixed to very small values. In practice,
we have not found any difference between using one condition or the
other. Both types of conditions give the maximum accretion rates of
mass and momentum (Benensohn et al., 1997).
For the outermost radial boundary we have, again, imposed different
conditions depending on the upstream or downstream position of the flow with
respect to the hole. In the upstream part, the initial asymptotic values
for all variables have been used at every time step. For the
downstream part, we have performed a linear extrapolation. This boundary
condition turns out to be the most critical one as, if the Mach number of
the flow is not high enough, unphysical information can be reflected
back
to the computational domain. This could be the reason why we did not
succeed in evolving low Mach number models. In future work we plan to use
characteristic conditions in this boundary to see if they work better.
Finally, we do not need to impose boundary conditions
on the angular direction as no boundaries exist.

We have tried different angular resolutions to analyze the convergence of the
numerical scheme.
We chose model MB2 of Table 1 for this purpose, as it showed signs of
unstable patterns in the wake.
This resolution study was motivated
by the recent work by Benensohn et al. (1997) where the
unstable patterns only appeared if the number of angular zones
to describe the entire circle was fine enough (roughly of the order of
200 zones). We performed 4 runs with 40, 80, 160 and 320 angular zones.
In all cases the initial conditions were identical and the flow was completely
uniform and totally symmetric with respect to the $\phi=0$ line.
With 40 zones the flow started stable and symmetric but after some time,
due to purely numerical reasons, mainly accumulated roundoff error,
an unstable pattern developed which even broke the wake. 
This was also visible with 80 angular zones as we will
show below. However, the numerical instability appeared later and was
much less violent (meaning that the solution was converging). 
With 160 and 320 zones we have not found any difference (even looking
at the specific numbers in the data files) between the two regions
above and below $\phi=0$.
For practical purposes we have always used a grid of 80 angular zones as
most of the models showed good late time behaviour with this resolution.

\section{Results}

\subsection{Flow morphology}

The evolution of the models listed in Table 1 is discussed here. 
In Figs. 2 and 3 we
show the final state of the first six models. We plot 
in Fig. 2 isocontours of
the logarithmic of the density normalized to the asymptotic value
while Fig. 3 shows isocontours of the total
velocity of the flow. We note in both figures
that all models present two clearly defined different regions
upstream and downstream the accreting hole. In the upstream part of the
flow, the isocontours are almost perfectly rounded, indicating
slight deviation from
spherical accretion. This is clearly demonstrated in Fig. 4 where
we plot a radial cut along the line passing through the points
$\phi=0$ and $\pi$ for several quantities and
only for model MB3 (the rest of models show a completely similar
behaviour). We can not directly compare the curves
for the density and total
velocity with the analytic values expected for spherical accretion
(see, e.g., Hawley, Smarr and Wilson 1984):
\begin{equation}
\rho(r) = {k\over {r^2\sqrt{{2M\over r}}}}
\end{equation}
\begin{equation}
v(r) = \sqrt{{2M\over r}}
\end{equation}
where $k$ is a constant. The reason is that 
we are assuming a fixed value for the velocity at the edge of the 
grid. For zero velocity at infinity this value is the free-fall
velocity at the given radius (assuming that this radius is smaller
than the transonic radius). However, the velocity pattern is clearly
spherical as, just by simply adding a radial term to the exact velocity, of,
e.g., the form $(r-r_{min})^{1/3}/a$, where $a$ is a constant, it is
possibly to get the numerical spherical solution.
It is worth to mention,
however, that the maximum values for the velocity at the innermost point
for all models completely agree with the expected free-fall values as can
be read off Table 2.
From Fig. 4 one can also notice that, at the rear part of the hole, the
density has always its maximum values, noticeably larger than the
values at the front part. This means that the majority of the material
is accreted onto the hole precisely at its rear part.
The velocity plot reaches a zero value at the
{\it stagnation point} behind the hole. All material inside 
the radius delimited by this point
is ultimately captured by the hole.

On the other hand, downstream the accretor, all models show the
presence of a well-defined tail shock that, starting very
close to the axis, propagates all the way up to the outer boundary of
the domain. This shock wave starts at the axis 
as a consequence of the gravitational bending that the matter feels,
which originates the collision and piling-up of the upper and lower
parts of the flow. Then, 
after developing the Bondi-Hoyle accretion {\it column}, the shock
starts to continuosly open to wider angles (due to the large pressure
gradient between the inner and outer parts of the wake) until it
finally acommodates itself in a fixed position.
The particular models that we have
considered are always characterized by a tail-shock. This shock never
propagates to the upstream part of the flow. As we can see in both 
figures the wake appears to be completely symmetric with respect to
the $\phi=0$ line for all models. As mentioned earlier, only slight
asymmetries develop for $\gamma=4/3$ models (MB2 and MB3). 
This is purely a numerical instability.
Results with larger resolutions (160 or 320 angular
zones), not shown here, reveal a fully converged and symmetric pattern.
However, with inadequate resolutions, 
this numerical instability can trigger asymmetric patterns which can,
eventually, broke the wake. These asymmetries
are more clearly noticeable in the velocity contours of Fig. 3.

The final model in Table 1, model UA2, is plotted in Figs. 5 and 6.
Fig. 5 shows isocontours of density and velocity while Fig. 6 shows
radial cuts along the line connecting $\phi=0$ and $\phi=\pi$. 
This model differs from the rest in having
a larger value for the asymptotic sound speed. This value is very close to 
its maximum given by $\sqrt{\gamma-1}$, $\gamma$ being the adiabatic
index of the fluid. 
Hence, this is an ultrarelativistic
model, both from the thermodynamic and kinematic points of view (note
that the asymptotic initial velocity is also very close to one).
The isocontour plots reveal a completely similar flow pattern to those of
Figs. 2 and 3. Note that
the steady-state solution is achieved here at much earlier times.
Clearly there is a correlation between flow velocity and stability: the
larger the speed, the more stable the pattern becomes (and the larger
the size of the accretor is). Let us also mention
that model UA2 is a large accretor, with $r_{min}=2.12 r_a$,
which can help in obtaining extremely stable patterns. The stability of
the large speed flows has also been confirmed with accretion rate plots
as we will show below.

In Table 2 we present the main quantitative results concerning the
evolution of these models. Comparing the morphology of the flow with the
axisymmetric configurations of Paper I we notice that the shock
opening angles are now smaller and, somehow closer to the expected
analytic values at large distances from the object
\begin{equation}
\theta_a = \sin^{-1}{1\over {\cal M}_{\infty}}.
\end{equation}
\noindent
We can also compute this analytic estimate using the
{\it relativistic definition} for
the Mach number (K\"onigl, 1980)
\begin{equation}
{\cal M}^{R}_{\infty} = {W \over W_s} {\cal M}_{\infty}
\end{equation}
\noindent
where $W$ is the Lorentz factor of the flow
and $W_s$ is the Lorentz-like factor for the sound speed, i.e.,
$W_s=(1-c^2_s)^{-1/2}$, with the sound speed defined as 
$c^2_s=\gamma p / \rho h$. With this definition, the expected values
for the {\it mildly} relativistic models (MA2-MC3) are slightly
modified: from $11.5$ to $10$ for models labeled `2' and from
$5.7$ to $4.9$ for models labeled `3'.
The corresponding value for UA2, which has a larger
asymptotic initial velocity, is now much different changing from
$19.5$ to $7.4$. In either case, the agreement with the exact 
solution is not too good.
There should also be mentioned that all the computed angles are
quite similar despite the different Mach numbers. The dependence
of the shock opening angle with the asymptotic Mach number is
plotted in Fig. 7.

The position of the stagnation point presents some 
similitudes and some differences
with respect to the axisymmetric models. The first thing we
notice is that the position has very little dependence on the incident
Mach number, as can be seen comparing models labeled '2' with
models labeled '3'. On the other hand there is a clear dependence
with the adiabatic index of the flow. 
The position decreases as $\gamma$
increases. This behaviour was also found in the axisymmetric models.
Comparing with those models of Paper I we find now considerable smaller
values. All models that can be directly compared (MA2, MB2 and MC2) are
now roughly a factor 2 smaller. This is however not the case for model UA2.

\subsection{Accretion rates}

We have computed the accretion rates of mass, radial momentum (drag rate)
and angular momentum. The results are summarized in Table 2 and
in Figs. 8-13 . These
quantities are excellent indicators to see if the solution ever 
approaches to a steady-state. In particular, monitoring the angular momentum
rate during the whole simulation is interesting to see if any
transient accretion disc forms.

Following the same procedure of Paper I we now compute the mass accretion
rate in the equatorial plane according to
\begin{equation}
\dot{m}=-\int D v^{r} r^2 d\phi
\end{equation}
\noindent
To compute the radial and angular momentum rates we have 
followed Petrich et al.
(1989) defining the variation in time of the $i$ momentum as
\begin{equation}
\dot{p}^i = -\int_{\partial V} T^{ij} \sqrt{-g} d\Sigma_j +
             \int_V (source)^i \sqrt{-g} dV
\end{equation}
where the flux integral is evaluated along the surface $\partial V$
of a given volume $V$ and the integral including the source terms is
a full volume integral. In practice, 
if we have the outer boundary sufficiently
far out, one can neglect the second integral as the source terms go to
zero in flat space. However, as this is not the case in our finite numerical
domain, we have always added the contribution of the sources
in our results. In the above expresion, $T^{ij}$ are the spatial components
of the stress-energy tensor and $g$ is the determinant of the four metric.

After some algebra, the final expresions we use are
\begin{eqnarray}
\dot{p}^r  = -r^2_c \int_{0}^{2 \pi} [S^r v^r + 
 p (1-{2M \over r_c}) ] d\phi +
\nonumber \\
\int_{r_{min}}^{r_c} \int_{0}^{2 \pi} 
 [-{M\over\alpha}(\tau+D) + \alpha r(S_{\phi}v^{\phi} -
 2S_r v^r) ] dr d\phi 
\end{eqnarray}
\noindent
for the radial momentum rate, and
\begin{equation}
\dot{p}^{\phi} = -r^2_c \int_{0}^{2 \pi} S^{\phi} v^r d\phi +
\int_{r_{min}}^{r_c} \int_{0}^{2 \pi}
 S_{\phi}v^r({M\over\alpha}-2\alpha r) dr d\phi
\end{equation} 
\noindent
for the angular momentum rate. In these expresions, $r_c$ is the
critical radius at which we have computed the rates.
In practice, the mass accretion rate has been obtained at the accretion
radius for all models except for model UA2 where it was computed at
$5M$ (the accretion radius is inside the horizon in this case). 
The momentum rates have been calculated at three different parts
of the grid just to check the influence of the source terms integral. 
These three radii are $r_{min}$, $r_a$ and $r_{max}$. We only plot
the results corresponding to the accretion radius.
All mass accretion rates have been 
normalized to the canonical value proposed
by Petrich et al. (1989). 
We can, thus, compare our results for the mass accretion
rates with those previously found in axisymmetric computations.
In addition, the radial momentum rates for models MA2-MC3 have
been scaled to one and the values corresponding to model UA2 are
displayed with respect to the same scale factor.

The evolution in time of the
different accretion rates appears in Figs. 8-13. Fig. 8 
shows the mass accretion rates for models MA2 to MC3 and Fig. 9 shows the
same quantity for model UA2. 
This mass rate has been scaled by the same canonical value used in Paper I.
The scaled radial momentum rates for models MA2 to MC3
are plotted in Fig. 10 while Fig. 11 shows this quantity for model UA2.
Finally, the angular momentum accretion rates for the different models appear
in Figs. 12 and 13.

From these figures we can again conclude that all models approach to a
remarkably well defined final steady-state. All the conclusions derived in
the axisymmetric simulations are still valid here. We find again that, given
a fixed value of ${\cal M}_{\infty}$, the mass accretion rate increases as 
$\gamma$ goes from 1.1 to $5/3$. The dependence with ${\cal M}_{\infty}$ for
a fixed value of $\gamma$ is, however, not too clear, and the numerical
values show only small discrepancies. 
This is particularly true for models MB2 and
MB3 which give the same final value. As in the axisymmetric 
simulations of Paper I,
the model with an almost maximum sound speed, model UA2, 
displays much larger values.
Comparing the values of Table 2 for the mass accretion rates with the
axisymmetric results,
we found that, for models MA2, MB2 and MC2 now we get roughly 10 times
larger values. For model UA2 the difference is even bigger, a factor 16.

The radial momentum accretion rate plots clearly show the same behaviour than
the mass accretion plots. After some initial time the solution relaxes to a
constant value. This happens after $t \approx 150M$ for all models
with low sound speed (MA2-MC3) and much earlier, $t \approx 35M$, for model
UA2. As mentioned previously, we have computed the radial momentum rate at three different locations.
Although we do not plot it here, we have verified that, as expected,
for larger values of the radial coordinate, the integral containing the source
terms in equation (18) becomes less relevant and
the solution converges to a constant value, both in radius and time. 

It is also interesting to look at the mass and radial momentum accretion
rates plots to compare the stability of the different models. 
First thing to notice is that both types
of figures display the same behaviour. Focussing on models MA2-MC3, we can
see that those with $\gamma=1.1$ are the most smooth and stable ones. As
$\gamma$ increases to $4/3$ there are some small and irregular oscillations.
Finally, for $\gamma=5/3$, the oscillations around the central equilibrium
value are clearly of constant period and their amplitude is fairly small.
We have also found that 
models MB2 and MB3 with higher resolution (160 angular zones) show the
same periodic feature than models MC2 and MC3. On the other hand, our
ultrarelativistic model, UA2, shows, apart from an oscillatory phase around
$110M$, a very smooth constant pattern. This is due to the large
velocities and large size of this accretor, and is also in agreement with
the rest of $\gamma=1.1$ models (MA2 and MA3).

Finally, we have computed the angular momentum accretion rates. By looking
at Figs. 12 and 13 one can clearly notice that the flow never
succeeds in getting high angular momentum values, except for some large
amplitude oscillations at late times for model MB3 and at an intermediate
time for model UA2. 
This means that no transient accretion discs
actually form, which would have not been the case if any flip-flop instability
would have developed. 
All models show, basically, the same behaviour: an initial
constant, almost zero value, 
followed by a period of oscillation around that central
value. In some cases, as model MC3, this oscillatory phase shows signs of
growing linearly, although at the time the simulation was stopped, the
amplitude of the oscillations
was still small. In particular, models MA2, MA3, MC2 and MC3 display very small
maximum amplitudes. Models with $\gamma=4/3$, MB2 and MB3, show,
however, larger oscillations and, in particular, model MB3 shows
a clear deviation from zero at very late times. 
As mentioned previously, the reason for this is again
purely numeric and directly related to the unresolved wake patterns obtained
with our canonical grid size.
The corresponding plot for model UA2 appears in Fig. 13. Although at some point
of the evolution it gets some large oscillations, it eventually settles down
to small values.

\section{Conclusions}

We have performed a detailed numerical study of relativistic Bondi-Hoyle
accretion onto a moving Schwarzshild black hole. We have extended our
previous work to include non-axisymmetric configurations. This has been
motivated by the possibility of finding unstable accretion patterns in the
flow. We have only considered flows which are initially uniform. We have
found that, contrary to Newtonian simulations, the non-axisymmetric
solution is always stable to tangetial oscillations. 
However, this statement should be relaxed, as the corresponding initial
setup in our relativistic wind simulations can not reproduce, for physical
reasons, the one used in classical simulations. In particular, we have 
considered values of the accretion radius which, when used in the classical
regime, would have not given rise to any instability as well. This may
indicate that, in the relativistic regime, a uniform flow 
accreting onto a black hole could always
remain symmetric and stationary. It might be worthwhile to go to much 
larger values of the asymptotic Mach number to see if any unstable
pattern actually develops.
These are, however, quite challenging simulations that would push the
code to its limits, as the asymptotic sound speed should be
considerably small.

We have checked the stability and stationarity of our
relativistic accretion patterns 
by analyzing the broad morphology of the flow with isocontour plots as
well as by integrating,
over all time of the simulation, global quantities as the
accretion rates of mass and momentum. 
In all cases considered, we have
found that the mass and radial momentum accretion rates 
approach asymptotically to very
accurate constant values in time. 
In addition, the angular momentum accretion rate
always remains in very low values, indicating the total absence of rotating
flows around the hole which could have triggered any tangetial
instability.

It maybe worthwhile to mention that, though we have restricted
in the present work to homogeneous flows, the code is capable of evolving
more complicated non-axisymmetric configurations where the flow has
some amount of angular momentum at infinity.
We are currently working in a parametric study of this kind of flows
(Pons et al., 1998).
In particular, we plan to gradually increase the
angular momentum of the incident
flow and study how it is transfered onto the hole via
alternate spiral shocks. At the same
time, we also plan to add rotation to the hole
-- a feature that the actual code used in this work already
incorporates --
extending thus the relativistic Bondi-Hoyle accretion to moving Kerr black
holes. Ultimately, we plan to perform these simulations in three spatial
dimensions.

\section*{Acknowledgments}

This work has been supported by the Spanish DGICYT (grant PB94-0973).
J.A.F has benefited from a European contract (nr. ERBFMBICT971902).
All computations have been performed at the AEI in Potsdam.

\newpage

\small

\begin{center}
    \begin{tabular}{|cccccccccc|}
    \multicolumn{10}{c}{\bf Table 1 \rm} \\
    \multicolumn{10}{c}{\bf Initial models\rm} \\
    \multicolumn{10}{c}{}\\
    \hline 
      MODEL & $c_{s_{\infty}}$ & $\gamma$ &
              ${\cal M}_{\infty}$ & 
              ${\cal M}^{R}_{\infty}$ & 
              $v_{\infty}$ &
              $r_a(M)$ & $r_{min} (r_a)$ & $r_{max} (r_a)$  &
              $t_f (M)$  \cr
\hline
      MA2     & $0.1$ & $1.1$  & $5.0$ & $5.74$ & $0.5$
              & $3.8$   & $0.57$  & $10.0$
              & $750$ \cr

      MA3     & $0.05$ & $1.1$  & $10.0$ & $11.53$ & $0.5$
              & $3.96$   & $0.55$  & $10.0$
              & $750$ \cr

      MB2     & $0.1$  & $4/3$ & $5.0$ & $5.74$ & $0.5$
              & $3.8$   & $0.57$  & $10.0$
              & $750$ \cr

      MB3     & $0.05$  & $4/3$ & $10.0$ & $11.53$ & $0.5$
              & $3.96$   & $0.55$  & $10.0$
              & $750$ \cr

      MC2     & $0.1$ & $5/3$  & $5.0$ & $5.74$ & $0.5$
              & $3.8$   & $0.57$  & $10.0$
              & $750$ \cr

      MC3     & $0.05$ & $5/3$  & $10.0$ & $11.53$ & $0.5$
              & $3.96$   & $0.55$  & $10.0$
              & $750$ \cr

      UA2     & $0.31$ & $1.1$    & $3.0$  & $7.76$ & $0.93$
              & $1.04$ & $2.12$  & $28.85$
              & $200$ \cr
\hline 
\multicolumn{10}{c}{}\\
\multicolumn{10}{c}{}\\
\end{tabular}

 $c_{s_{\infty}}$ is the asymptotic sound speed, $\gamma$ is the
 adiabatic exponent, ${\cal M}_{\infty}$ is the asymptotic Mach
 number (Newtonian definition), 
 ${\cal M}^{R}_{\infty}$  is the asymptotic relativistic Mach number (see  definition in the text),
 $v_{\infty}$ is the asymptotic flow velocity,
 $r_a$ is the accretion radius, $r_{min}$ and $r_{max}$ are the
 minimum and maximum radial values of the computational domain
 and $t_f$ is the final time, in units of $M$,
 at which the simulation is stopped.
 The specific names of the models are chosen to be in agreement
 with those of Paper I.

\vfill
\end{center}

\normalsize

\small

\begin{center}
    \begin{tabular}{|ccccccccccccc|}
    \multicolumn{13}{c}{\bf Table 2 \rm} \\
    \multicolumn{13}{c}{\bf Summary of results \rm} \\
    \multicolumn{13}{c}{}\\
    \hline 
      MODEL & $\theta_{a}$       & $\theta_c$         &
              ${\bar r}_{sp}$                         &
              $\rho_{max}^u$     & $\rho_{max}^d$     &
              $v_{max}^u$        & $v_{max}^d$        &
              ${\cal M}_{max}^u$ & ${\cal M}_{max}^d$ &
              ${\dot{m}}/{\dot{m}_{can}}$             &
              ${\dot{p}^r}$           &
              ${\dot{p}^{\phi}}$ \cr
\hline
      MA2 & $11.5$ & $10$   & $3.8$  & $12.7$ & $291.7$
          & $0.96$ & $0.92$ & $8.8$  & $4.6$  & $37.6$
          & $-0.38$ & $1\times 10^{-6}$ \cr

      MA3 & $5.7$  & $9$    & $3.5$  & $12.9$ & $351.4$
          & $0.96$ & $0.92$ & $19.9$ & $4.9$  & $35.6$
          & $-0.33$ & $1.4 \times 10^{-4}$ \cr

  MB2$^*$ & $11.5$ & $15$   & $2.6$  & $12.8$ & $106.4$
          & $0.96$ & $0.90$ & $6.8$  & $2.5$  & $43.2$
          & $-0.40$ & $1.6 \times 10^{-2}$ \cr

  MB3$^*$ & $5.7$  & $12$   & $2.5$  & $13.0$ & $109.9$
          & $0.97$ & $0.90$ & $24.8$ & $2.4$  & $43.3$
          & $-0.46$ & $5.5$ \cr

      MC2 & $11.5$ & $17$   & $1.5$  & $12.9$ & $47.9$
          & $0.96$ & $0.86$ & $4.6$  & $1.5$  & $106.9$
          & $-0.98$ & $3 \times 10^{-5}$ \cr

      MC3 & $5.7$  & $17$   & $1.6$  & $12.9$ & $53.5$
          & $0.96$ & $0.85$ & $13.5$ & $1.5$  & $109.5$
          & $-0.87$ & $5.5 \times 10^{-4}$ \cr

      UA2 & $19.5$ & $12$   & $5.6$  & $15.3$ & $284.0$
          & $0.99$ & $0.88$ & $3.2$  & $2.8$  & $410.5$
          & $-140.1$ & $7.5$ \cr
\hline 
\multicolumn{9}{c}{}\\
\multicolumn{9}{c}{}\\
\end{tabular}
The asterisk in models MB2 and MB3 indicates that the results have been
obtained with $160$ angular zones.
$\theta_a$ is the analytical value for the shock opening angle at large
distances according to
$\sin^{-1}{1\over{\cal M}_{\infty}}$ (Newtonian definition for
the Mach number), 
$\theta_c$ is the numerical
result, ${\bar r}_{sp}$ is the
mean value of the stagnation point position,
$\rho_{max}^u$ is the maximum in the density in the upstream direction
along the line $\phi=\pi$,
$\rho_{max}^d$ is the corresponding maximum in the density downstream
$\phi=0$,
$v_{max}^u$ is the maximum upwind velocity,
$v_{max}^d$ is the
maximum downwind velocity,
${\cal M}_{max}^u$ is the maximum value of the upwind Mach number and
${\cal M}_{max}^d$ is its corresponding maximum value in the
downwind direction. In addition ${\dot{m}}/{\dot{m}_{can}}$ is the
normalized mass accretion rate (to the canonical value used in
Paper I) and ${\dot{p}^r}$ and
${\dot{p}^{\phi}}$ are the scaled rates of radial and angular momentum,
respectively. For the latter quantity we only indicate the maximum values.

\vfill
\end{center}

\normalsize

\newpage

\begin{figure}
\centerline{\psfig{figure=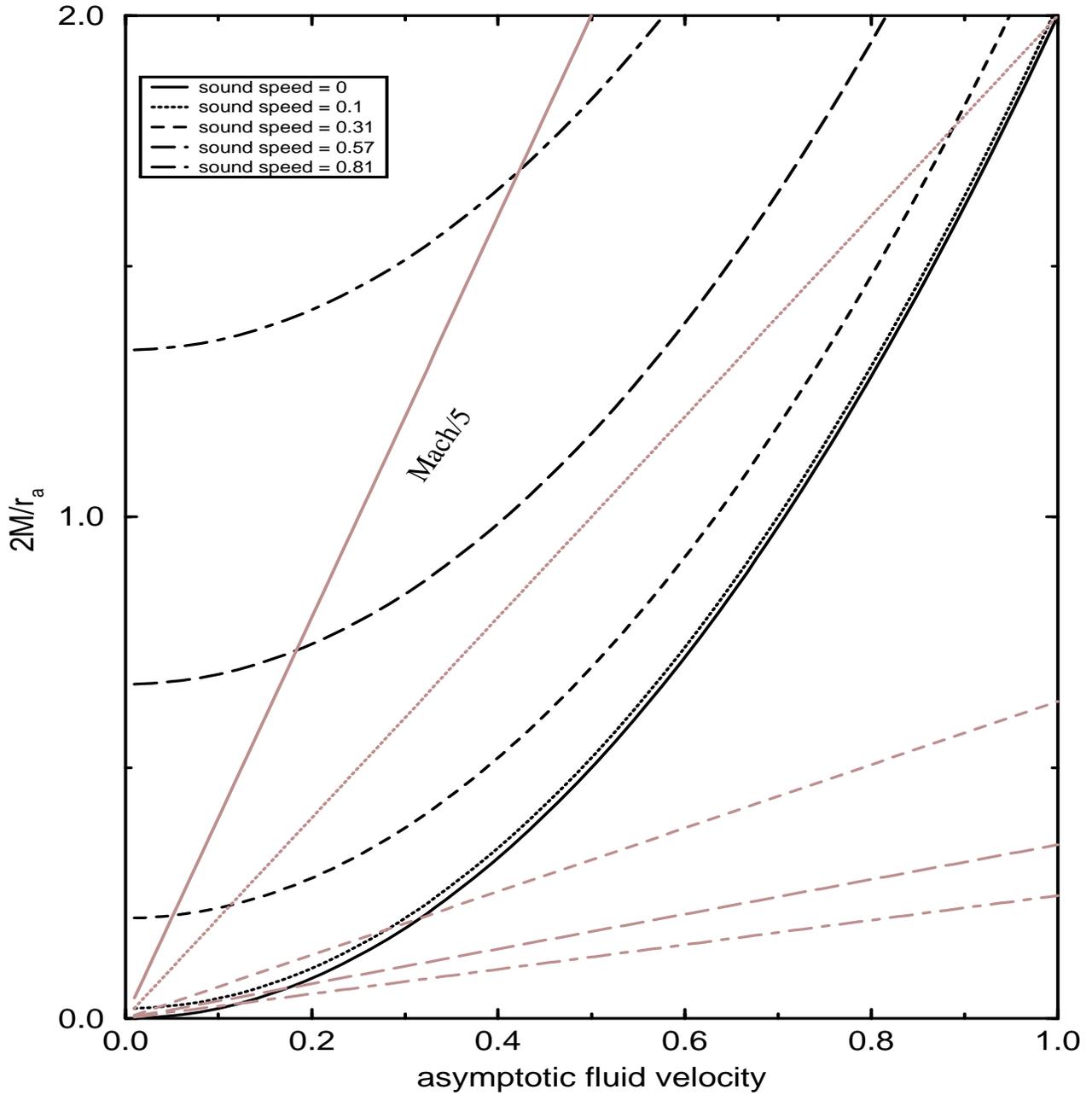,width=7.0in,height=7.0in}}
\caption{{ \protect \small Minimum size of the accretor in units of the
  accretion radius as a function of the asymptotic flow velocity. The curved
  lines are parametrized for different sound speed values,
  specifically, $0$
  ({\it solid}), $0.1$ ({\it dotted}), $0.31$ ({\it dashed}), $0.57$
  ({\it long dashed}) and $0.81$ ({\it dot-dashed}). In addition, the
  straight lines represent the asymptotic Mach number (over $5$) for the same
  set of sound speeds (except the solid line
  which corresponds to $c_{s_{\infty}}=0.05$).
}}
\label{fig1}
\end{figure}

\newpage

\begin{figure}
\centerline{\psfig{figure=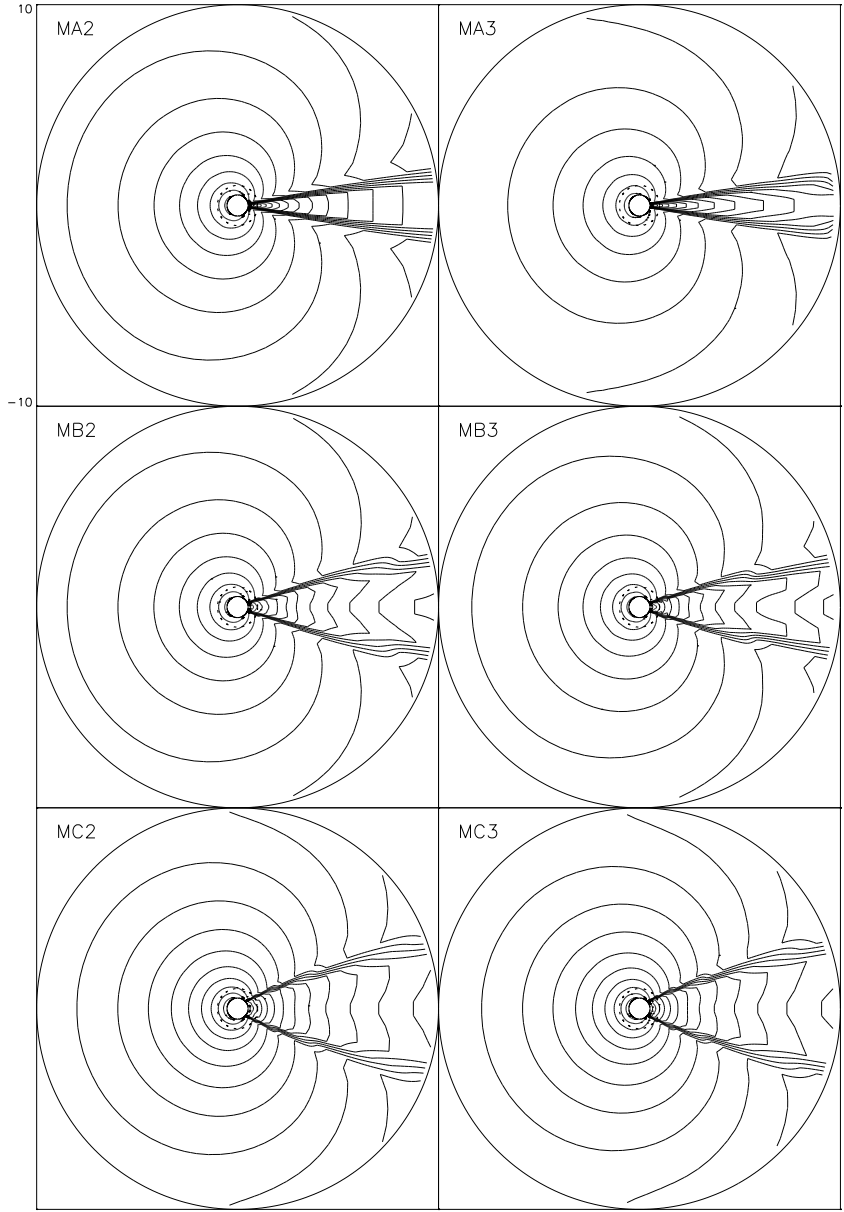,width=7.0in,height=7.0in}}
\caption{{ \protect \small 
Isocontours of the logarithm of the scaled
 density at the final time of the evolution ($t=750M$) for models
 MA2 to MC3. The dotted line indicates the position of the accretion radius.
 All models show a well-defined tail shock and a spherical upstream
 accretion pattern.
}}
\label{fig2}
\end{figure}

\newpage

\begin{figure}
\centerline{\psfig{figure=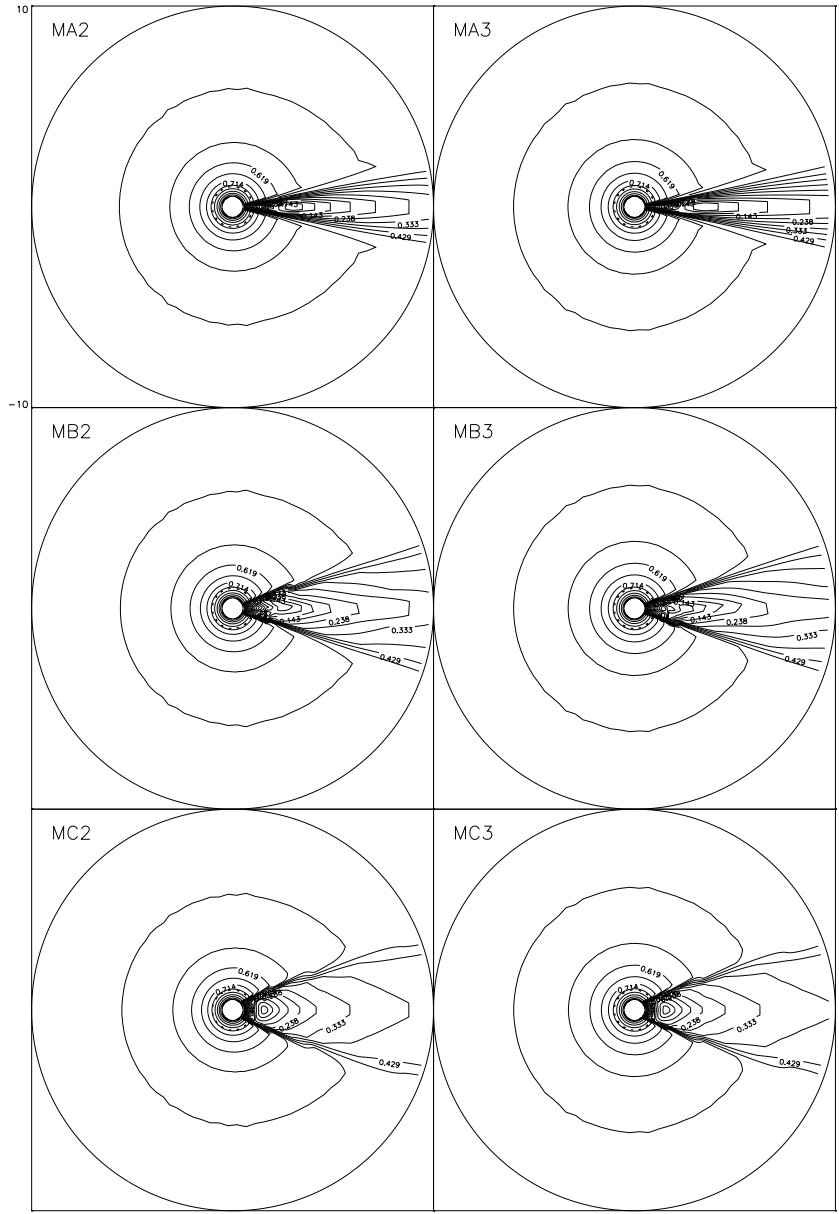,width=7.0in,height=7.0in}}
\caption{{ \protect \small 
Isocontours of the total velocity
 at the final time of the evolution ($t=750M$) for models
 MA2 to MC3.
}}
\label{fig3}
\end{figure}

\newpage

\begin{figure}
\centerline{\psfig{figure=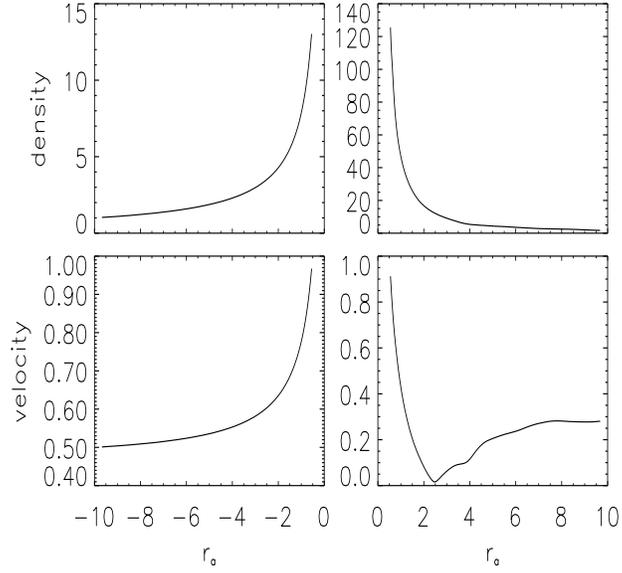,width=4.0in,height=4.0in}}
\caption{{ \protect \small 
 Radial plot of density and velocity
 for Model MB3. The left panels show the upstream part of the
 flow while the right panels show the downstream part.
 The upstream solution clearly indicates spherical flow plus a constant
 velocity added.
}}
\label{fig4}
\end{figure}

\begin{figure}
\centerline{\psfig{figure=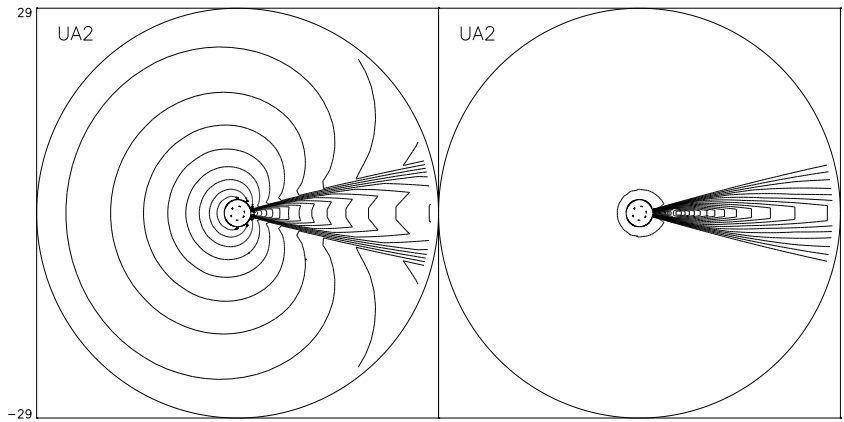,width=7.0in,height=2.6in}}
\caption{{ \protect \small 
Isocontours of the
 logarithm of the scaled density (left) and total velocity (right)
 at the final time of the evolution ($t=200M$) for model UA2. The
 accretion radius, doted line, is inside the horizon.
}}
\label{fig5}
\end{figure}

\newpage

\begin{figure}
\centerline{\psfig{figure=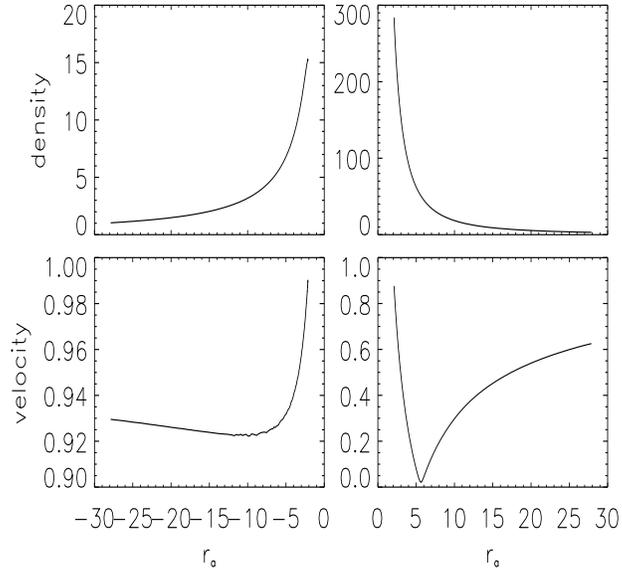,width=4.0in,height=4.0in}}
\caption{{ \protect \small 
Radial plot of density and velocity
 for Model UA2. The left panels show the upstream part of the
 flow while the right panels show the downstream part.
}}
\label{fig6}
\end{figure}

\begin{figure}
\centerline{\psfig{figure=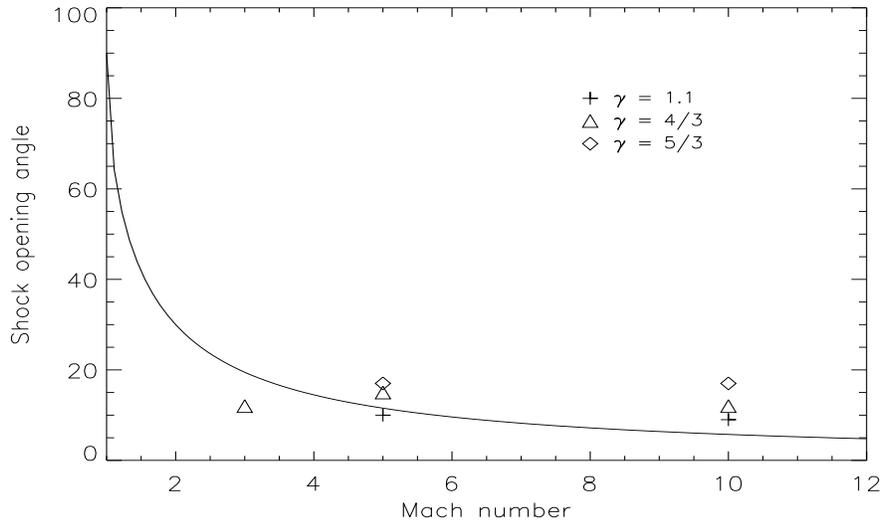,width=5.0in,height=3.0in}}
\caption{{ \protect \small 
 Shock opening angle versus asymptotic Mach number. The solid line
 represents the analytic values at large distance from the hole. The symbols
 indicate the numerical estimations for the different models.
}}
\label{fig7}
\end{figure}

\newpage

\begin{figure}
\centerline{\psfig{figure=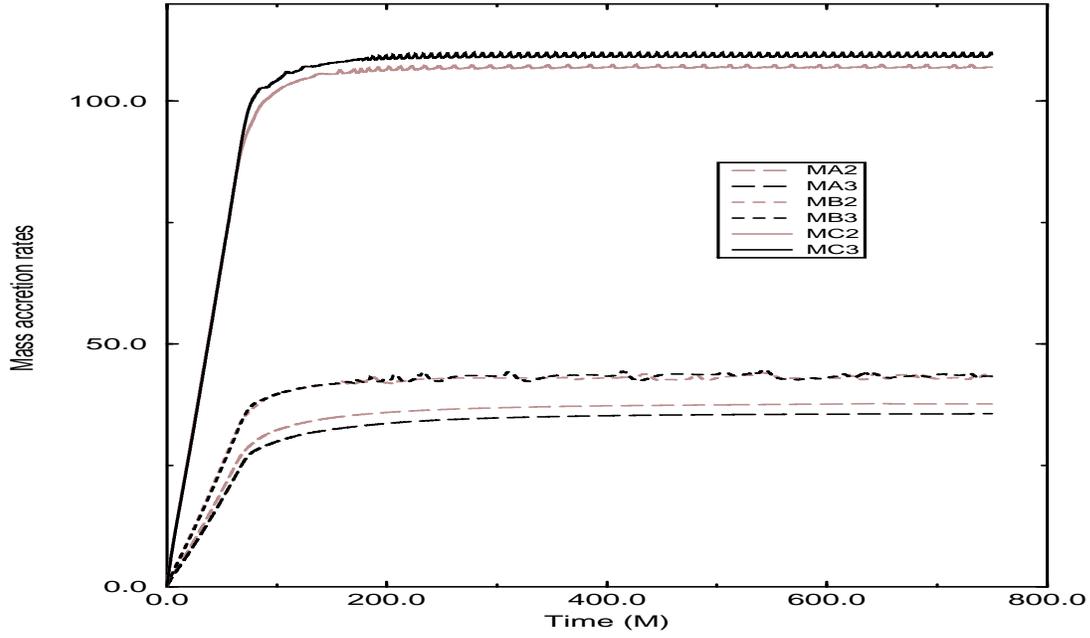,width=6.0in,height=3.5in}}
\caption{{ \protect \small 
Normalized mass accretion rate evolution for models MA2-MC3.
}}
\label{fig8}
\end{figure}

\begin{figure}
\centerline{\psfig{figure=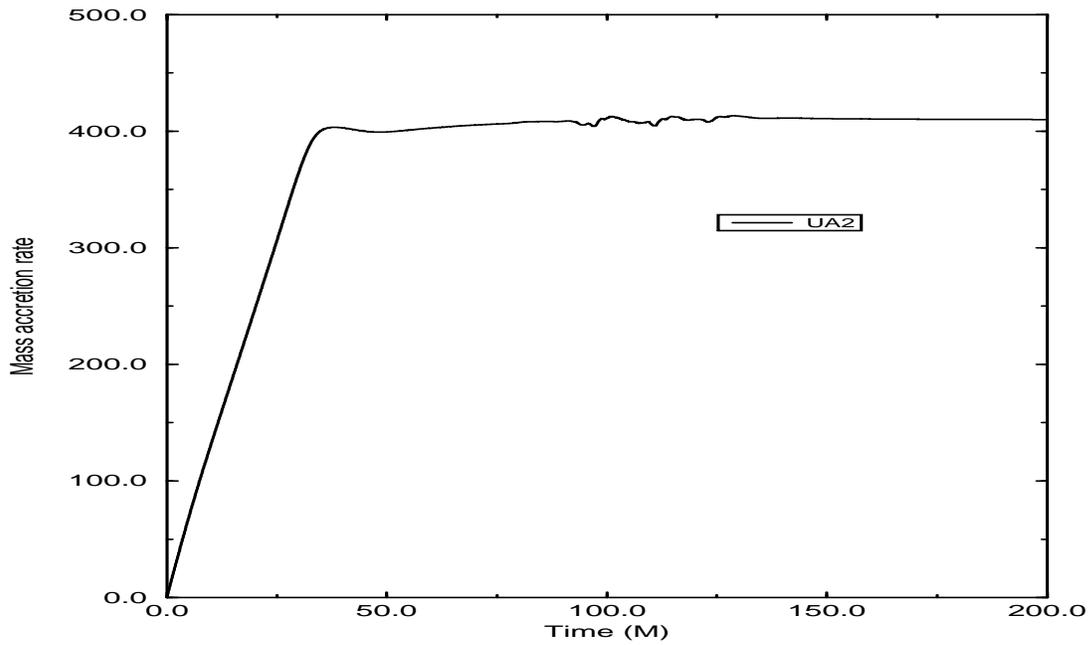,width=6.0in,height=3.5in}}
\caption{{ \protect \small 
Normalized mass accretion rate evolution for model UA2.
}}
\label{fig9}
\end{figure}

\newpage

\begin{figure}
\centerline{\psfig{figure=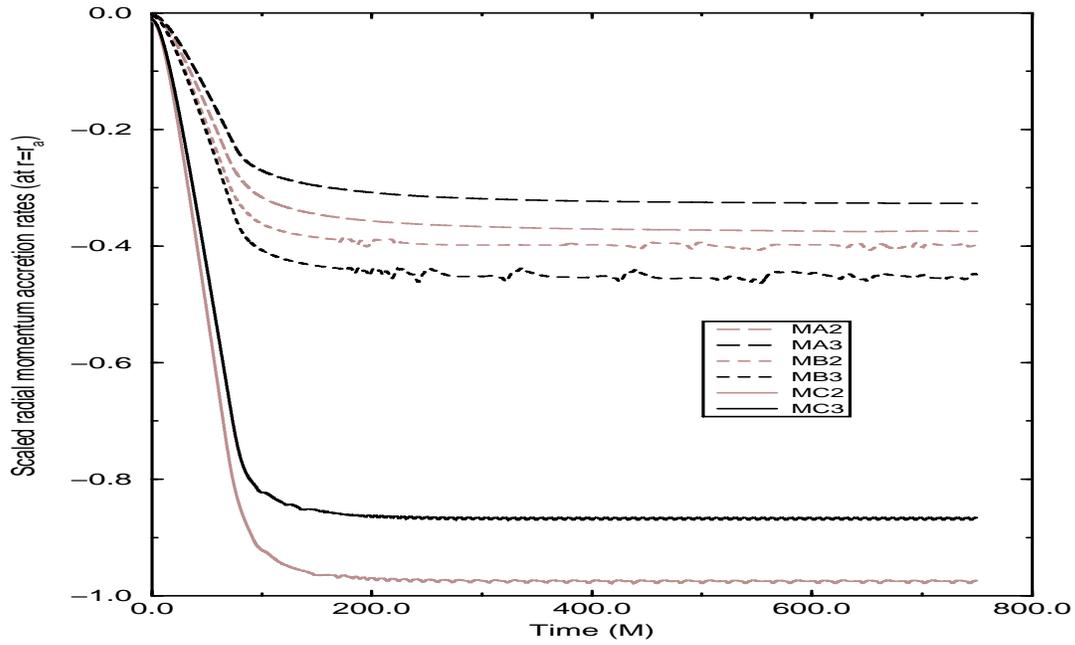,width=6.0in,height=3.5in}}
\caption{{ \protect \small 
Normalized radial momentum accretion rate evolution for models MA2-MC3.
}}
\label{fig10}
\end{figure}

\begin{figure}
\centerline{\psfig{figure=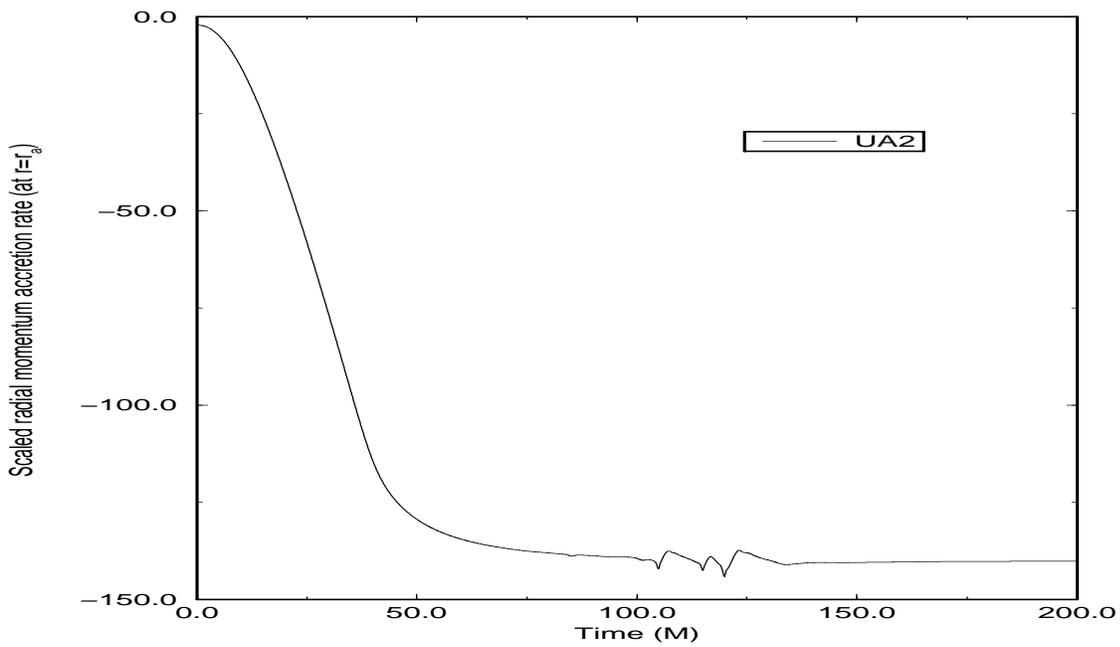,width=6.0in,height=3.5in}}
\caption{{ \protect \small 
Normalized radial momentum accretion rate evolution for model UA2.
}}
\label{fig11}
\end{figure}

\newpage

\begin{figure}
\centerline{\psfig{figure=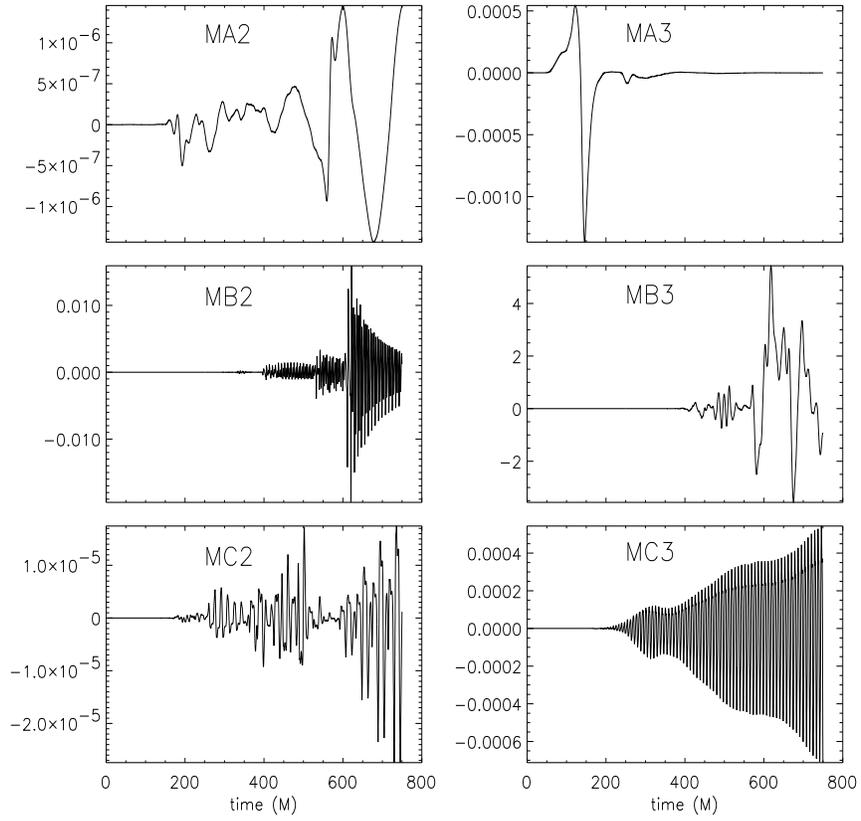,width=4.7in,height=4.7in}}
\caption{{ \protect \small 
Angular momentum accretion rate evolution for models MA2-MC3.
}}
\label{fig12}
\end{figure}

\begin{figure}
\centerline{\psfig{figure=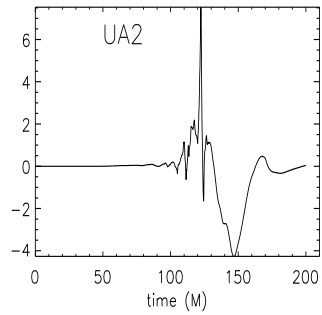,width=2.in,height=2.in}}
\caption{{ \protect \small 
Angular momentum accretion rate evolution for model UA2.
}}
\label{fig13}
\end{figure}

\end{document}